\documentclass[preprint]{article}
\usepackage{graphicx}% Include Figure files
\usepackage{dcolumn}% Align table columns on decimal point
\usepackage{bm}% bold math
\usepackage{amsmath}
\usepackage{amssymb}
\usepackage{multirow}
\usepackage{caption}
\usepackage{epstopdf}
\usepackage{hyperref}
\usepackage{cite}

\begin{document}
{\setlength{\oddsidemargin}{1.2in}
\setlength{\evensidemargin}{1.2in} } \baselineskip 0.55cm
\begin{center}
{\LARGE {\bf Compact stars with gravitational wave echoes in $f(R,L_{m},T)$ gravitational thoery }}
\end{center}
\date{\today}
\begin{center}
  Meghanil Sinha, S. Surendra Singh \\
Department of Mathematics, National Institute of Technology Manipur,\\
Imphal-795004,India\\
Email:{ meghanil1729@gmail.com, ssuren.mu@gmail.com}\\
 \end{center}
 
\textbf{Abstract}: This work explores the gravitational wave echoes (GWEs) from the compact stellar configurations in the backdrop of $f(R,L_{m},T)$ gravity within static and spherically symmetric framework. Our study has utilized the MIT Bag model and color-favour-locked (CFL) phase equations of state (EoS) for matter description. Mass-radius profiles were determined by solving the hydrostatic equilibrium equations. Model parameter variations were used to assess the configuration stability here. TOV solutions helped to evaluate compactness. Our results indicate that MIT bag model and CFL EoS in $f(R,L_{m},T)$ modified gravitational theory are capable of producing GWEs. The calculated wave frequencies lie within the range of $ 7.5-11 $ kHz range. We have also demonstrated that how different gravitational theory parametrization within $f(R,L_{m},T)$ theory affect our star structure and echo frequency characteristics. Surface redshift and adiabatic index analysis confirm the stability of our stellar model here.\\

 
\section{Introduction}\label{sec1} 

\hspace{0.8cm} Over a century old, Einstein's gravity theory continues to illuminate our understanding of the Universe while undergoing extensive experimental validation. Predicting Mercury's perihelion precession accurately lists as one of the significant validation, while others include gravitational wave (GW) detections evolving from black holes (BH) regimes and neutron star coalescence recorded by LIGO ( Laser Interferometer Gravitational-Wave Observatory ), VIRGO collaboration in conjunction with the Event Horizon Telescope (EHT)'s first BH shadow image \cite{N1,N3}. Consequently, General Relativity (GR)'s predictions align closely with the observational evidence gathered mainly over the past century.\\
However, from a modern astrophysical and cosmological perspective, we inhabit an age of accelerating Universe expansion and grapple with the mystery of dark matter (DM) \cite{R1,R3}. This accelerated expansion can be attributed to two main hypotheses. One explanation suggests the requirement of a hypothetical dark energy (DE) to account for the current expansion \cite{R4,R5}. In the second approach, extending GR's geometric sector through modified gravity theories offers a potential replacement of DE, which allows for the understanding of the current cosmic acceleration without relying on the mysterious DE \cite{R5,R6,R7}. Likewise, the mystery of the missing mass has been addressed by proposing DM, a non-luminous and non-baryonic form of matter \cite{R7}. From the viewpoint of alternating modified gravitational theories this can be evaded. Moreover, a range of observational and theoretical findings encourages investigation into alternative gravity theories \cite{R10,R11}. Today, modified gravity theories are considered a promising direction for exploring physics beyond GR \cite{R12,R13}.\\
The strong gravity near the compact celestial objects (neutron stars, etc.) and BH event horizons' presents testing grounds and poses notable challenges \cite{N22,N23}. Compact celestial objects' extreme mass and density forge unique platforms where gravity exhibits extraordinary intensity. These environments provide a unique opportunity for empirical validation of gravitational physics. Compact star's characteristics such as mass, radius, compactness, redshift and adiabatic index provide valuable insights. Direct astronomical observations can yield mass and radius making them vital parameters. These values reveal information about the framework of the star and aid in determining the EoS that describes its internal composition. New advanced telescopes with enhanced sensitivity allow for more accurate results. Advanced technology facilitates testing GR and its extensions in strong gravitational fields.\\
In the recent times, from observational evidence from pulsars, X-ray bursts, NICER ( Neutron Star Interior Composition Explorer ) suggests the possible presence of exotic states of matter, like the deconfined quarks \cite{D31,D32} deep within the dense stellar cores. Initially the hypothesis was proposed that compact objects might be composed of strange quark matter \cite{D33,D35}. These stars were categorized as distinct class of strange quark matter which includes a nearly equal mix of up, down and strange quarks with a small electron component for neutrality. The MIT bag model offers a powerful way to understand strange quark matter \cite{D36}. The bag model features a spherical region of free quark movement. The MIT bag model was used to investigate strange and quark stars in the context of GR \cite{D37,D41} and also in alternative gravitational theories \cite{D44,Q,T}. Quark stars serve as a probe for the phase transition from hadronic to quark matter in dense interior of the star, advancing our understanding of quantum chromodynamics (QCD). In high-density regimes, QCD's ground state is believed to be the CFL phase. A paired phase of quarks could exist in the neutron star cores. Quark pairing modifies the quasiparticle spectrum, affecting the transport properties substantially. CFL matter, characterized by strong interactions is predicted to become stable at sufficiently high densities \cite{D48}. While useful, the MIT bag model falls short in capturing the complexities of strongly interacting quark systems. QCD-driven research has led to modified quark star models incorporating the CFL phase \cite{D44}. Quarks may form various color superconducting phase under extreme density conditions \cite{D49,D51}.\\
The study of such compact objects has intriguing implications for gravitational wave detection, where future research may validate different compact matter models, deepening our understanding of stellar objects. The compactness of these stars would enable them to produce GWEs \cite{RT42,RT44,RT46}. The concept of GWEs from compact stars were originated in the study of \cite{RT43}. Research has showed MIT bag model EoS in GR can produce GWEs \cite{RT47}. Studies have investigated and analyzed GWEs from compact stellar models in GR, considering different EoS \cite{RT48,RT50}. Modified gravitational theories such as $f(R)$ and $f(R,T)$ have been studied for GWEs across multiple models \cite{R,RT}. GWEs may provide mean to test various alternative gravity theories. Inspired by the previous research, this study investigates the GWEs from compact stars in $f(R,L_{m},T)$ modified gravitational theory. Our research focuses on the model of  $ f(R,L_{m},T) = R + \gamma T L_{m} $ characterised by the constant $ \gamma $\cite{M}. Our $f(R,L_{m},T)$ gravitational theory encompasses both $f(R,T)$ and $f(R,L_{m})$  gravity models providing a unified framework. This theory posits the gravitational Lagrangian as a function of the Ricci scalar, energy momentum trace and matter density. The role of $ \gamma $ in shaping the internal properties of compact stars and GWEs from them has been analysed. Our work employs the MIT bag and CFL phase EoS for modeling strange quark matters. The research also includes the stability study of our constructed models.\\
This paper is constructed in the following manner : Section (\ref{sec1}) provides a brief introduction while section (\ref{sec2}) outlines the $f(R,L_{m},T)$ gravity theory. Section (\ref{sec3}) discusses the EoSs for the compact matter composition, while the GWEs are examined in section (\ref{sec4}), including the calculated results for our model. The physical viability of our stellar model are being demonstrated in section (\ref{sec5}) followed by the results and the final discussions are presented in section (\ref{sec6}). Throughout the article, we have adopted the geometrized units $c=G=1$.\\

\section{$f(R,L_{m},T)$ gravity formalism}\label{sec2} 

\hspace{0.8cm}The $f(R,L_{m},T)$ gravity model proposed here merges $f(R,T)$ and $f(R,L_{m})$  with gravitational Lagrangian as a function of $R$(Ricci scalar), $T$(energy-momentum trace) and $L_{m}$(matter Lagrangian) \cite{N13}. The action for the gravitational framework is expressed as\\
\begin{equation}\label{1}
\texttt{S} = \frac{1}{16 \pi}\int f(R,L_{m},T)\sqrt{-g}d^{4}x + \int L_{m}\sqrt{-g}d^{4}x.
\end{equation}\\
Here $g$ symbolizes the determinant of the metric tensor. We get the field equation by taking the variation of the action w.r.t $g^{\eta\chi}$ as,\\
\begin{eqnarray}\label{2}
&& f_{R}(R,L_{m},T)R_{\eta\chi} - \frac{1}{2}[f(R,L_{m},T) - (f_{Lm}(R,L_{m},T) + 2f_{T}(R,L_{m},T)L_{m})L_{m}]g_{\eta\chi} + 
\nonumber \\
&& (g_{\eta\chi}\Box-\nabla_{\eta}\nabla_{\chi})f_{R}(R,L_{m},T) = [8 \pi +\frac{1}{2}(f_{Lm}(R,L_{m},T) + 2f_{T}(R,L_{m},T)L_{m})]T_{\eta\chi} + f_{T}\Pi_{\eta\chi}.
\end{eqnarray}\\
Here $ \Box = \partial_{\eta}(\sqrt{-g}g^{\eta\chi}\partial_{\chi})/\sqrt{-g}, f_{R}(R,L_{m},T) = \frac{\partial f(R,L_{m},T)}{\partial R}, f_{T}(R,L_{m},T) = \frac{\partial f(R,L_{m},T)}{\partial T}, f_{L_{m}}(R,L_{m},T) = \frac{\partial f(R,L_{m},T)}{\partial L_{m}}, R_{\eta \chi} = $Ricci tensor, $ \nabla_{\eta} $= covariant derivative w.r.t symmetric connection to $g_{\eta\chi}$. The tensor $ \Pi_{\eta\chi} $ is specified as \cite{N26},\\
\begin{equation}\label{3}
\Pi_{\eta\chi} = 2g^{\psi\nu}\frac{\partial^L_{m}}{\partial g^{\eta\chi} \partial g^{\psi\nu}}.
\end{equation}\\
Setting $ f(R,L_{m},T) = f(R) $ leads to the field equation in $f(R)$ gravity \cite{N27}. For $ f(R,L_{m},T) = f(R,T) $ it gives $ f(R,T) $ gravity equations while for $ f(R,L_{m},T) = f(R,L_{m}) $ it yields $ f(R,L_{m}) $ theory. $ f(R,L_{m},T) = R $ gives the usual field equations for pure GR as,\\
\begin{equation}\label{4}
R_{\eta\chi} - \frac{1}{2} g_{\eta\chi}R = 8 \pi T_{\eta\chi}.
\end{equation}\\
The covariant divergence of equation (2) results in the non-conservation equation for the energy momentum tensor as\\
\begin{eqnarray}\label{5}
\nabla^{\eta}T_{\eta\chi} &=& \frac{1}{8\pi + [f_{T}(R,L_{m},T) + \frac{1}{2}f_{L_{m}}(R,L_{m},T)]}\times
\nonumber \\
&& \Big(\nabla_{\chi}(L_{m}((f_{T}(R,L_{m},T) + \frac{1}{2}f_{L_{m}}(R,L_{m},T)) - T_{\eta\chi}\nabla^{\eta}(f_{T}(R,L_{m},T) + \frac{1}{2}f_{L_{m}}(R,L_{m},T))
\nonumber \\
&& \nabla^{\eta}(f_{T}\Pi_{\eta\chi}) - \frac{1}{2}(f_{T}(R,L_{m},T)\nabla_{\chi}T + f_{L_{m}}(R,L_{m},T)\nabla_{\chi}L_{m})))\Big).
\end{eqnarray}\\
Here we have used $ \nabla^{\eta}R_{\eta\chi} = \nabla_{\chi}\frac{R}{2} $ with $ (\Box\nabla_{\chi} - \nabla_{\chi}\Box) \xi = R_{\eta\chi}\nabla^{\eta}\xi $ valid for any scalar field $ \xi $. Taking the trace, results in a second order differential equation as\\
\begin{eqnarray}\label{6}
&& 3\Box f_{R}(R,L_{m},T) + Rf_{R}(R,L_{m},T) - 2[f(R,L_{m},T) - 2(f_{T}(R,L_{m},T) + \frac{1}{2}f_{L_{m}}(R,L_{m},T))L_{m}] = 
\nonumber \\
&& [8 \pi + (f_{T}(R,L_{m},T) + \frac{1}{2}f_{L_{m}}(R,L_{m},T))]T + f_{T}(R,L_{m},T)\Pi.
\end{eqnarray}\\
Here $ \Pi = $ trace of $ \Pi_{\eta\chi} $. For $ f(R,L_{m},T) = f(R) $, this becomes the familiar dynamical equation for $ f(R) $ gravitational theory \cite{N29}. Non-linear term in $ R $ might lead to a non-zero scalar curvature in the exterior. We have considered here thus $ f(R,L_{m},T) = R + \gamma TL_{m} $ with $ \gamma $ as matter-geometry coupling constant here. In this case, equations (2) and (5) simplify to\\
\begin{equation}\label{7}
G_{\eta\chi} = [ 8 \pi + \frac{3\gamma}{2}\gamma(T+2L_{m})]T_{\eta\chi} + \gamma L_{m}(\Pi_{\eta\chi} + L_{m}g_{\eta\chi})
\end{equation}\\
\begin{eqnarray}\label{8}
\nabla^{\eta}T_{\eta\chi} &=& \frac{\gamma}{8 \pi + \gamma(L_{m} + \frac{T}{2})}\times \Big(\nabla_{\chi}(L_{m}^{2} + \frac{1}{2}TL_{m})
\nonumber \\
&& T_{\eta\chi}\nabla^{\eta}(L_{m} + \frac{T}{2}) - \nabla^{\eta}(L_{m}\Pi_{\eta\chi}) - \frac{1}{2}(L_{m}\nabla_{\chi}T + T\nabla_{\chi}L_{m})\Big).
\end{eqnarray}\\
Here $ G_{\eta\chi} = $ Einstein tensor. For $ \gamma = 0 $ leads to the recovery of Einstein field equation and the conservation equation for GR. For compact stars being composed of adiabatic and isotropic fluids, the energy-momentum tensor of the perfect fluid is considered as\\
\begin{equation}\label{9}
T_{\eta\chi} = (\rho + p)\upsilon_{\eta}\upsilon_{\chi} + pg_{\eta\chi}
\end{equation}\\
with $ \rho = $ density, $ p = $ pressure of the isotropic fluid and $ \upsilon_{\eta} $ representing the four-velocity vector. The static and spherically symmetric metric for compact stars is expressed as\\
\begin{equation}\label{10}
ds^{2} = -e^{w(r)} dt^{2} + e^{x(r)} dr^{2} + r^{2}(d\theta^{2} + \sin^{2}\theta d\phi^{2}) 
\end{equation}\\
where $ w(r) $ and $ x(r) $ are the two unknown functions, both of which depend on $ r $(the radial co-ordinate). The choice $ L_{m} = -\rho $ is made here, considering its relevance for the compact stellar model. In line with conventional GR, energy density directly contributes to the gravitational source, with matter density driving the gravity field. $ L_{m} = p $ would result in varied mass-radius relations and stability outcomes as demonstrated in \cite{A7}. We choose here $ L_{m} = -\rho $ in accordance to compact object treatment and simplify numerical computations. This ensures matter's intuitive gravitational effects with $ \rho $ as the main source. Hence, we get from equation (7),\\
\begin{equation}\label{11}
G_{\eta\chi} = [8 \pi + \frac{3\gamma}{2}(p - \rho)]T_{\eta\chi} - \gamma \rho^{2}g_{\eta\chi}. 
\end{equation}\\
Considering the spherically symmetric metric, the $ 00 $ and $ 11 $ components result in\\
\begin{equation}\label{12}
e^{-x}(\frac{x'}{r} - \frac{1}{r^{2}}) + \frac{1}{r^{2}} = 8 \pi \rho + \frac{\gamma}{2}(3p - \rho)\rho
\end{equation}\\
\begin{equation}\label{13}
e^{-x}(\frac{w'}{r} + \frac{1}{r^{2}}) - \frac{1}{r^{2}} = 8 \pi p + \frac{3\gamma}{2}(p - \rho)p - \gamma \rho^{2}.
\end{equation}\\
In the non-conservation equation of the energy-momentum tensor in equation (8), we obtain\\
\begin{equation}\label{14}
\frac{dp}{dr} + \frac{1}{2}\frac{dw}{dr}( \rho + p ) = \frac{\gamma[4 \rho \rho' + 3p(\rho' - p')]}{16 \pi + 3\gamma(p - \rho)}.
\end{equation}\\
Prior to discussing the compact star's internal composition, the mass function is defined via\\
\begin{equation}\label{15}
e^{w(r)} = e^{-x(r)} = 1 - \frac{2m}{r}
\end{equation}\\
and consequently the TOV equations appear as\\
\begin{equation}\label{16}
\frac{dm}{dr} = 4 \pi r^{2} \rho + \frac{\gamma r^{2}}{4}( 3p - \rho)\rho
\end{equation}\\
\begin{equation}\label{17}
\frac{dp}{dr} = -\frac{(\rho + p)[ 4 \pi rp + \frac{m}{r^{2}} + \frac{3\gamma r}{4}(p - \rho)p - \frac{\gamma r}{2} \rho^{2}]}{(1-\frac{2m}{r})[1 + \frac{\gamma(3p( 1 - \frac{d \rho}{dp}) - 4\rho (\frac{d \rho}{dp}))}{16 \pi + 3\gamma( p - \rho )}]}.
\end{equation}\\
The TOV equations are to be solved from centre $(r=0)$ to the surface at $(r=R)$ for this choice meeting the boundary conditions 
\begin{equation}\label{18}
m(0) = 0 \hspace{0.5cm} \rho(0) = \rho_{centre}  \hspace{0.5cm} p(0) = p_{centre} 
\end{equation}\\
with the central values of pressure $ p_{centre}  $ and density $ \rho_{centre} $. The pressure drops to $ 0 $ at the stellar surface. In this approach, the gravitational mass comes out to be $ m = \emph{M} $( at the surface ). The solution at the interior meets the vacuum exterior at the surface. Thus, at the surface the metric is described as $ e^{w(r)} = e^{-x(r)} = 1 - \frac{2\emph{M}}{R} $.\\

\section{Equation of State}\label{sec3}

\hspace{0.8cm}With an EoS relating the energy density and fluid pressure, the TOV equations can give us the physical features of the compact stars. We will take the help of the MIT bag model initially to characterize the dense matter here with the form \cite{RT37}\\
\begin{equation}\label{19}
\rho = 3p + 4 \verb"B". 
\end{equation}\\
where $ \verb"B" $ is the bag constant. This is the EoS for the deconfined quark matter (up, down and strange quarks) with confinement pressure via the bag constant. We also consider the strange quark matter in the CFL phase, described by its EoS \cite{RT38}. A completely up, down and strange quark-made ultra relativistic star is thought to be absolutely stable. In conditions of large density with lower temperature, quark matter is a color superconductor with quark Fermi gas with Cooper pair condensate at the Fermi surface \cite{D96}. Quark matter can shift to varying phases depending on the quark pairing patterns \cite{D97}. At sufficiently high density, quark matter converts to CFL state \cite{D44}. In this state, quarks pair into Cooper pairs of different colors and flavors. The EoS for such may involve somewhat more complex non-linear analytical relation as \cite{D49}\\
\begin{equation}\label{20}
\rho = 3p + 4\verb"B" - \frac{9 \sum \beta^{2}}{\pi^{2}}
\end{equation}\\
$ \beta^{2} $ and $ \sum $  are represented as\\
\begin{equation}\label{21}
\beta^{2} = -3\sum + (9\sum^{2} + \frac{4}{3}\pi^{2}(\verb"B" + P))^{\frac{1}{2}}
\end{equation}\\
and\\
\begin{equation}\label{22}
\sum = \frac{-m_{s}^{2}}{6} + \frac{2\triangle^{2}}{3}
\end{equation}\\
with the pairing gap denoted by $ \triangle $ and strange quark mass as $ m_{s} $. In this work we have looked into the massless quarks $( m_{s} = 0 )$ and $( m_{s} = 100 MeV )$ \cite{RT64}.\\

\section{Gravitational wave echoes}\label{sec4}

\begin{figure}[ht!]
\centering
\includegraphics[scale=0.5]{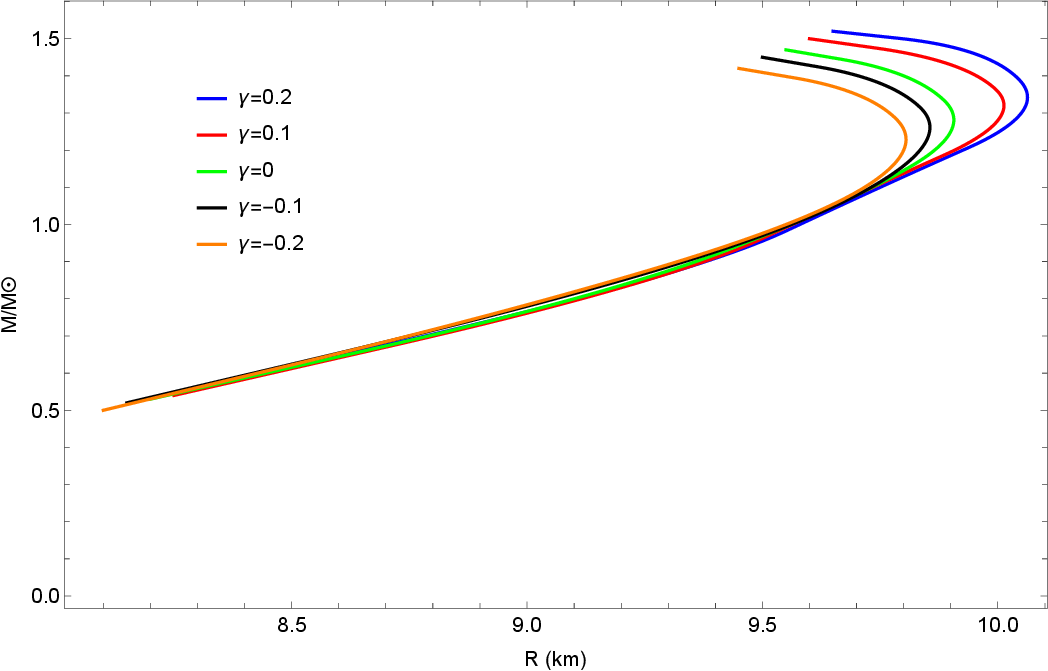}
\caption{M-R plot for the CFL phase state, $ m_{s} = 0, \triangle = 350MeV $}\label{3a}
\end{figure}

\begin{figure}[ht!]
\centering
\includegraphics[scale=0.5]{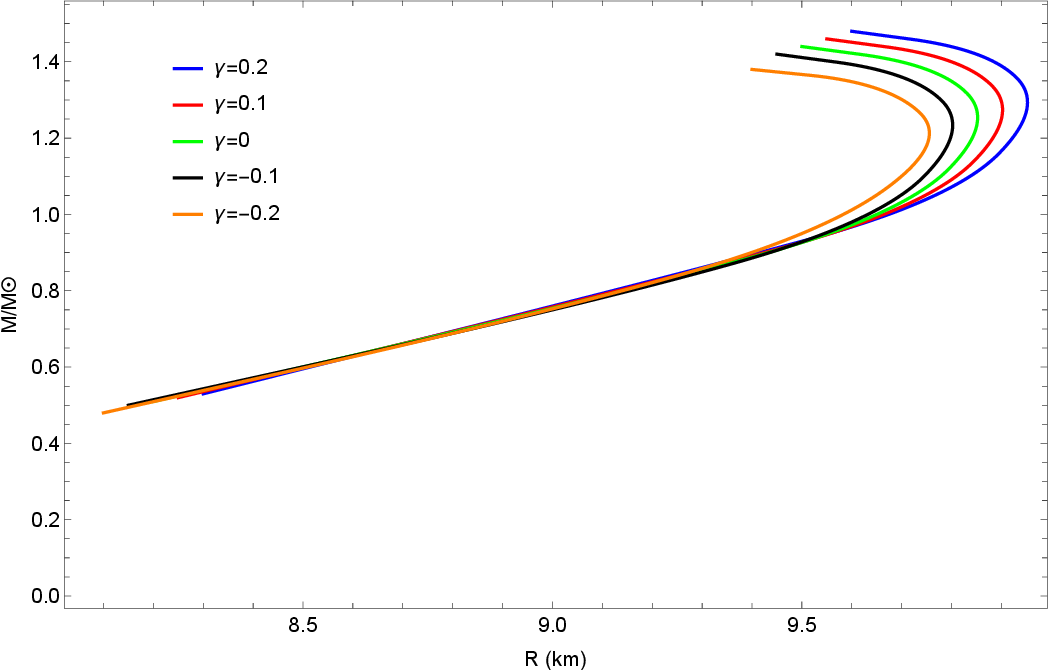}
\caption{M-R plot for the MIT Bag model , $ m_{s} = 100 MeV, \triangle = 350MeV $}\label{3b}
\end{figure}

\hspace{0.8cm}The observation of GWs gave a new direction to the understanding of the BHs and compact stars. Detection of merging binary BHs prompted researchers to look into exotic compact objects behaving like BH mimics. Unlike BHs, these compact structures can have high compactness without forming an event horizon. High compactness led to the proposals of such objects generating GWEs off their potential barrier \cite{RT43}. However, the existence of such compact objects doesn't preclude them developing rotational instabilities. Ultra compact stars may suffer from ergoregion instability along with non-linear instabilities \cite{RT63}. Linear and non-linear stability can be disturbed by very long lived modes in such scenarios. GWs from a distant merger hitting the surface get reflected at the photon sphere, leading to multiple reflections and refractions after a delay. For echoes of GWs to be generated, a photon sphere at $ R_{PH}= 3\emph{M} $(total mass $ \emph{M} $) is needed for such objects. Without event horizons, unlike BHs, compact stellar structures need a radius greater than the Buchdahl radius $ R_{BU} = \frac{9}{4} \emph{M} $. It is worth pointing out that the applicability of the Buchdahl limit is within GR \cite{RT72}, for modified gravitational theories it is counted as $ R_{BU} = \frac{9 \emph{M}}{4-\frac{3c}{2}} $, where $ c = 4 \pi \rho_{R} (R) R^{2} $, $ \rho_{R} = $ pressure at the boundary $ r = R $. In the works of \cite{RT73}, the value $ c $ is addressed as the negative. To get echo frequencies, we at first focus on the echo times given as \\
\begin{equation}\label{23}
\tau_{E} = \int_{0}^{3\emph{M}} e^{\frac{x(r)-w(r)}{2}} dr.
\end{equation}\\
Deriving the metric functions from equation (\ref{15}), we can have the echo frequencies as\\
\begin{equation}\label{24}
\omega_{E} \thickapprox \frac{\pi}{\tau_{E}}.
\end{equation}\\
\begin{table}[h!]
\centering
\caption{ List of some possible compact stars }
\begin{tabular}{||p{5.5cm}|p{2.9cm}|p{2.0cm}|p{2.0cm}||}
\hline\hline
\hspace{1cm}$ Compact \hspace{0.2cm} star $ & \hspace{1cm}$ M(M_{\bigodot}) $ & \hspace{0.6cm}$ R $(km) & \hspace{0.4cm}$ Ref $ \\
\hline\hline
$\hspace{1.1cm} PSR-B0943 + 10 $ & \hspace{1cm}$ 0.02 $ & $ 2.6 $ &  \cite{R113} \\[1pt]
\hline
$\hspace{1.1cm} HER X-1 $ & \hspace{0.7cm}$ 0.85\pm0.15 $ & $ 8.1\pm0.41 $ &  \cite{R97} \\[1pt]
\hline
$\hspace{1.1cm} 4U1538-52 $ & \hspace{0.7cm}$ 0.87\pm0.07 $ & $ 7.866\pm0.21 $ &  \cite{R112} \\[1pt]
\hline
$\hspace{1.1cm} SMC X-1 $ & \hspace{0.7cm}$ 1.04\pm0.09 $ & $ 8.301\pm0.2 $ &  \cite{R84} \\[1pt]
\hline
$\hspace{1.1cm} 4U 1728-34 $ & \hspace{1cm}$ 1.1 $ & $ 9 $ &  \cite{R112} \\[1pt]
\hline
$\hspace{1.1cm} SAX J1748.9-2021 $ & \hspace{0.7cm}$ 1.33\pm0.33 $ & $ 10.93\pm2.09 $ & \cite{R110}\\[1pt]
\hline
$\hspace{1.1cm} PSR 1937+21 $ & \hspace{1cm}$ 1.4 $ & $ 6.6 $ & \cite{R108}\\[1pt]
\hline
$\hspace{1.1cm} Cyg X-2 $ & \hspace{0.7cm}$ 1.44\pm0.06 $ & $ 9.0\pm0.5 $ & \cite{R107}\\[1pt]
\hline
$\hspace{1.1cm} CEN X-3 $ & \hspace{0.7cm}$ 1.49\pm0.08 $ & $ 9.178\pm0.13 $ & \cite{R97}\\[1pt]
\hline
$\hspace{1.1cm} 4U 1820-30 $ & \hspace{0.7cm}$ 1.58\pm0.06 $ & $ 9.1\pm0.4 $ & \cite{R106}\\[1pt]
\hline
$\hspace{1.1cm} SAX J1808.4-3658 $ & \hspace{1cm}$ 1.6 $ & $ 11 $ & \cite{R100}\\[1pt]
\hline
$\hspace{1.1cm} 4U 1822-37 $ & \hspace{0.7cm}$ 1.69\pm0.13 $ & $ 10 $ & \cite{R102}\\[1pt]
\hline
$\hspace{1.1cm} 4U 1608-52 $ & \hspace{0.7cm}$ 1.74\pm0.14 $ & $ 9.3\pm1.0 $ & \cite{R100}\\[1pt]
\hline
$\hspace{1.1cm} PSR J1614-2230 $ & \hspace{0.7cm}$ 1.97\pm0.04 $ & $ 9.69\pm0.2 $ & \cite{R96}\\[1pt]
\hline
$\hspace{1.1cm} PSR J0348+0432 $ & \hspace{0.7cm}$ 2.01\pm0.04 $ & $ 12.605\pm0.35 $ & \cite{R95}\\[1pt]
\hline
$\hspace{1.1cm} 4U 1636-536 $ & \hspace{0.7cm}$ 2.02\pm0.12 $ & $ 9.6\pm0.6 $ & \cite{R94}\\[1pt]
\hline\hline
\end{tabular}
\label{Tab:1}
\end{table}\\
Figures (\ref{1}), (\ref{3a}) and (\ref{3b}) depict the TOV equation solutions for the compact stars in (MIT bag, CFL phase) for different values of the parameter. In these specific models and data set, the mass increases as the parameter increases. In other words, sufficiently large positive values of $ \gamma $ lead to an increase in both the maximum mass and the associated radius. A bag constant value of $ (168 MeV)^{4} $ is being used in this study here. With $ \gamma = 0 $ representing the GR case in these figures, we found that compactness decrease with increasing $ \gamma $ in the M-R curves. Varying $ \gamma $ leads to different stellar structure ( masses and radii ), GWEs from the stellar structures have different frequencies for the MIT bag and CFL EoS.\\
\begin{table}[h!]
\centering
\caption{ Strange stars with bag constant = $ (168 MeV)^{4} $ for the MIT bag model }
\begin{tabular}{||p{5.5cm}|p{2.9cm}|p{2.0cm}|p{2.0cm}||}
\hline\hline
\hspace{1cm}$ \gamma(10^{-79}s^{4}/kg^{2}) $ & \hspace{1cm}$ M(M_{\bigodot}) $ & \hspace{0.6cm}$ R $(km) & \hspace{0.2cm} GWE frequency(kHz)  \\
\hline\hline
$\hspace{1.1cm} -0.2 $ & \hspace{0.8cm}$ 1.893 $ & $ 10.91 $  &  $ 10.1 $ \\[1pt]
\hline
$\hspace{1.1cm} -0.1 $ & \hspace{0.8cm}$ 1.821 $ & $ 11.23 $  &   $ 9.4 $  \\[1pt]
\hline
$\hspace{1.5cm} 0 $ & \hspace{0.8cm}$ 1.745 $ & $ 11.57 $  &   $ 8.7 $  \\[1pt]
\hline 
$\hspace{1.3cm} 0.1 $ & \hspace{0.8cm}$ 1.667 $ & $ 11.8 $  &   $ 8.1 $  \\[1pt]
\hline
$\hspace{1.3cm} 0.2 $ & \hspace{0.8cm}$ 1.589 $ & $ 12.13 $  &   $ 7.5 $  \\[1pt]
\hline\hline
\end{tabular}
\label{Tab:2}
\end{table}\\
\begin{table}[h!]
\centering
\caption{ Strange stars with CFL phase $( m_{s} = 0, \hspace{0.1cm} \triangle = 350MeV )$ }
\begin{tabular}{||p{5.5cm}|p{2.9cm}|p{2.0cm}|p{2.0cm}||}
\hline\hline
\hspace{1cm}$ \gamma(10^{-79}s^{4}/kg^{2}) $ & \hspace{1cm}$ M(M_{\bigodot}) $ & \hspace{0.6cm}$ R $(km) & \hspace{0.2cm} GWE frequency(kHz)  \\
\hline\hline
$\hspace{1.1cm} -0.2 $ & \hspace{0.8cm}$ 2.03 $ & $ 10.41 $  &  $ 10.8 $ \\[1pt]
\hline
$\hspace{1.1cm} -0.1 $ & \hspace{0.8cm}$ 1.957 $ & $ 10.73 $  &   $ 10.1 $  \\[1pt]
\hline
$\hspace{1.5cm} 0 $ & \hspace{0.8cm}$ 1.883 $ & $ 11.07 $  &   $ 9.5 $  \\[1pt]
\hline 
$\hspace{1.3cm} 0.1 $ & \hspace{0.8cm}$ 1.794 $ & $ 11.37 $  &   $ 8.8 $  \\[1pt]
\hline
$\hspace{1.3cm} 0.2 $ & \hspace{0.8cm}$ 1.713 $ & $ 11.65 $  &   $ 8.2 $  \\[1pt]
\hline\hline
\end{tabular}
\label{Tab:3}
\end{table}\\
\begin{table}[h!]
\centering
\caption{ Strange stars with CFL phase $( m_{s} = 100MeV, \hspace{0.1cm} \triangle = 350MeV )$ }
\begin{tabular}{||p{5.5cm}|p{2.9cm}|p{2.0cm}|p{2.0cm}||}
\hline\hline
\hspace{1cm}$ \gamma(10^{-79}s^{4}/kg^{2}) $ & \hspace{1cm}$ M(M_{\bigodot}) $ & \hspace{0.6cm}$ R $(km) & \hspace{0.2cm} GWE frequency(kHz)  \\
\hline\hline
$\hspace{1.1cm} -0.2 $ & \hspace{0.8cm}$ 1.967 $ & $ 10.67 $  &  $ 10.2 $ \\[1pt]
\hline
$\hspace{1.1cm} -0.1 $ & \hspace{0.8cm}$ 1.884 $ & $ 10.93 $  &   $ 9.6 $  \\[1pt]
\hline
$\hspace{1.5cm} 0 $ & \hspace{0.8cm}$ 1.802 $ & $ 11.27 $  &   $ 9.0 $  \\[1pt]
\hline 
$\hspace{1.3cm} 0.1 $ & \hspace{0.8cm}$ 1.723 $ & $ 11.58 $  &   $ 8.4 $  \\[1pt]
\hline
$\hspace{1.3cm} 0.2 $ & \hspace{0.8cm}$ 1.645 $ & $ 11.87 $  &   $ 7.9 $  \\[1pt]
\hline\hline
\end{tabular}
\label{Tab:4}
\end{table}\\
Figures (\ref{2}) and (\ref{4}) visualize echo frequency variation with $ \gamma \epsilon [-0.2,0.2]\times 10^{-79}s^{4}/kg^{2} $ for MIT bag and CFL phase respectively. A nearly linear relationship is observed for the GWE frequency w.r.t $ \gamma $. The nature of the compactness is also evident from the figure (\ref{5}) for different EoS. We have drawn the graphs for the mass as a function of the radial co-ordinate for the MIT bag and CFL phase with ( $ m_{s} = 0 $ and $ m_{s} = 100 MeV $ ) in figures (\ref{6}), (\ref{7}) and (\ref{8}) respectively. With radially outward, it increases for different parameter values of $ \gamma $.\\
\begin{figure}[ht!]
\centering
\includegraphics[scale=0.5]{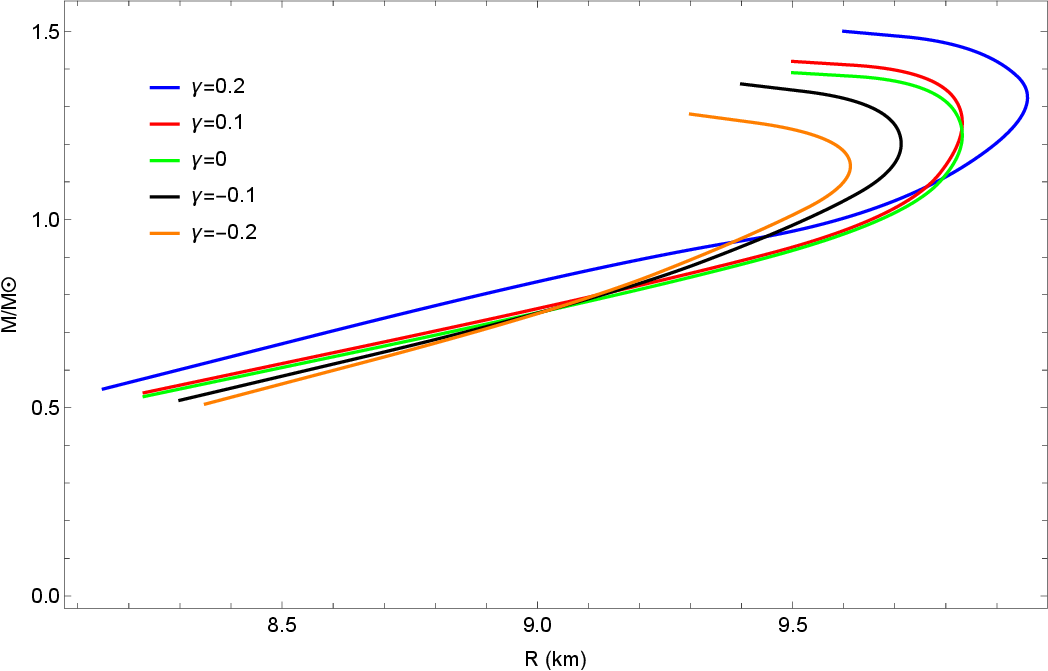}
\caption{M-R plot for the MIT Bag model , $ B = (168 MeV)^{4}, \gamma \epsilon [-0.2,0.2]\times 10^{-79}s^{4}/kg^{2}, \gamma = 0 $(GR) }\label{1}
\end{figure}\\
\begin{figure}[ht!]
\centering
\includegraphics[scale=0.5]{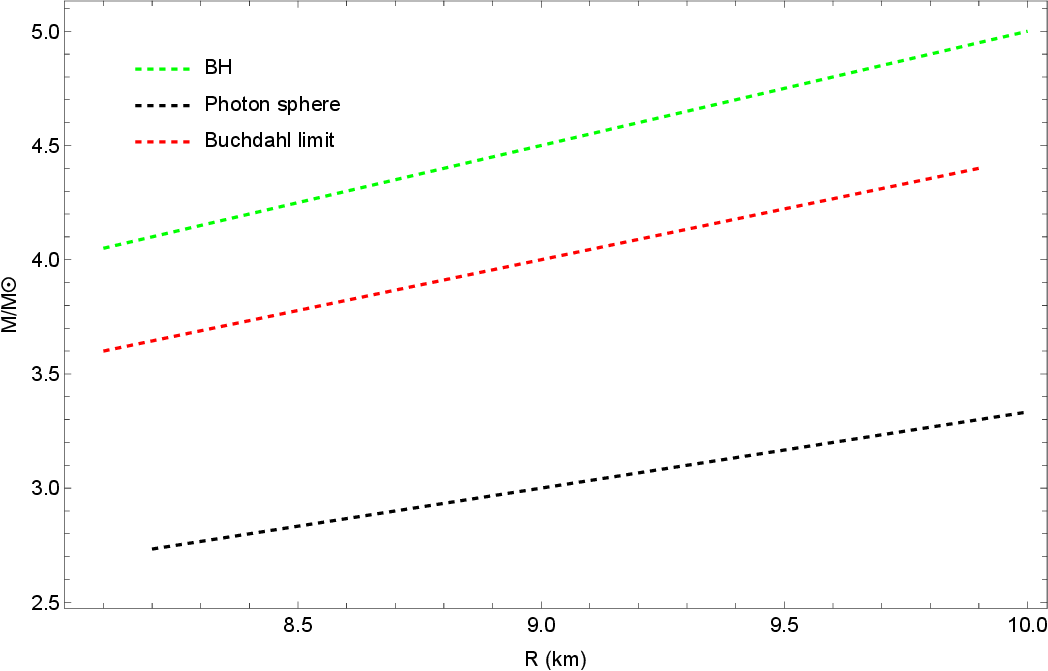}
\caption{ Plot for the photon sphere, Black hole and Buchdahl limit for our model }\label{1a}
\end{figure}\\
\begin{figure}[ht!]
\centering
\includegraphics[scale=0.5]{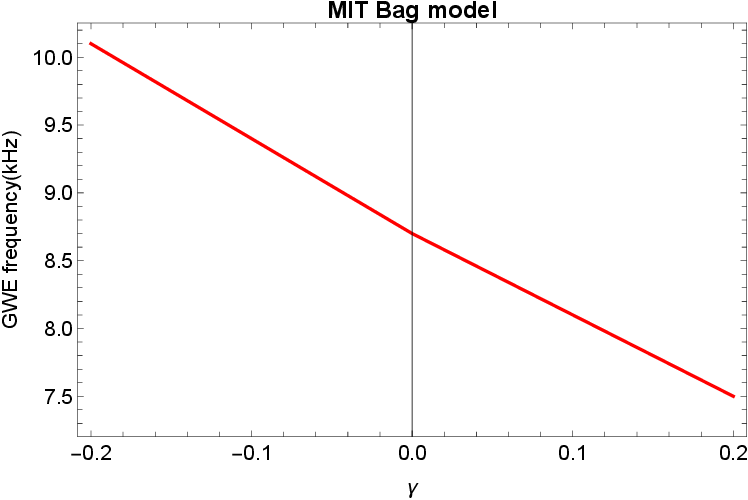}
\caption{Graph of GWE frequency within allowed $ \gamma $ for MIT bag model}\label{2}
\end{figure}\\
\\
\\
\begin{figure}[ht!]
\centering
\includegraphics[scale=0.5]{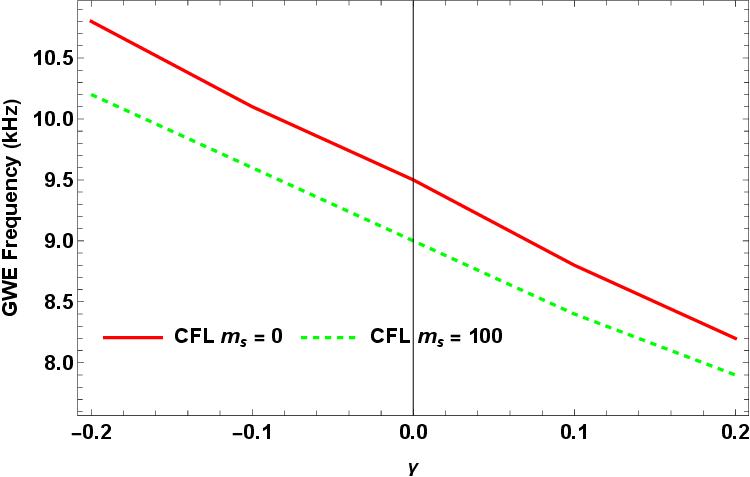}
\caption{Graph of GWE frequency for the CFL phases with $ \triangle = 350MeV $}\label{4}
\end{figure}\\
\begin{figure}[ht!]
\centering
\includegraphics[scale=0.5]{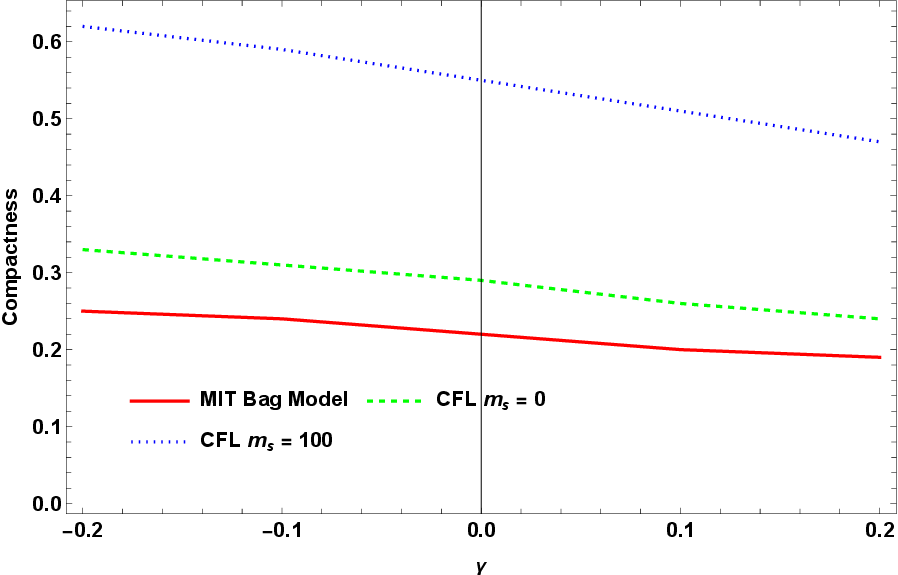}
\caption{Variation in the compactness of the stellar interior for MIT bag and CFL phase }\label{5}
\end{figure}\\
\begin{figure}[ht!]
\centering
\includegraphics[scale=0.5]{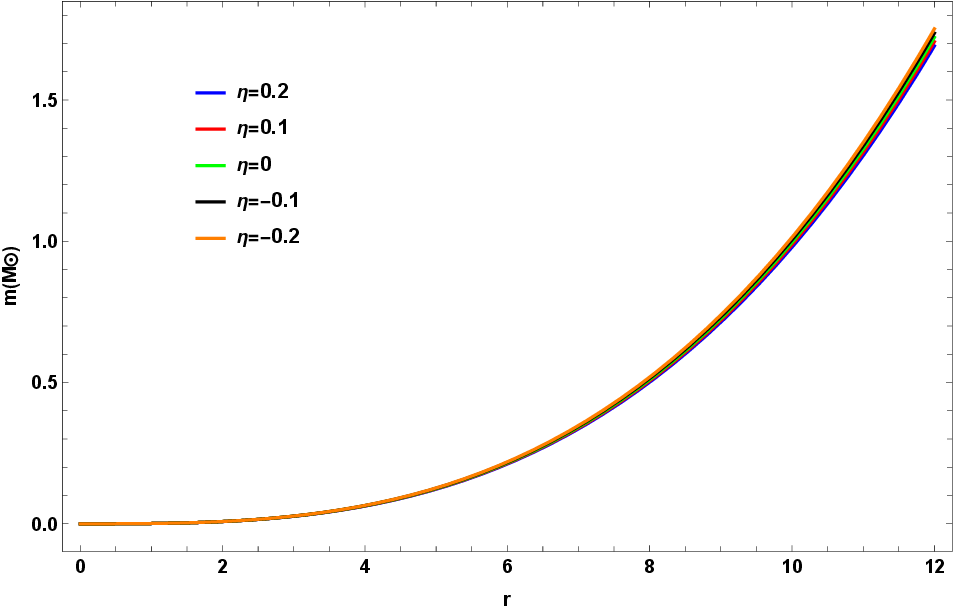}
\caption{Mass dependency on radial co-ordinate for MIT bag, $ B = (168 MeV)^{4} $ }\label{6}
\end{figure}\\
\begin{figure}[ht!]
\centering
\includegraphics[scale=0.5]{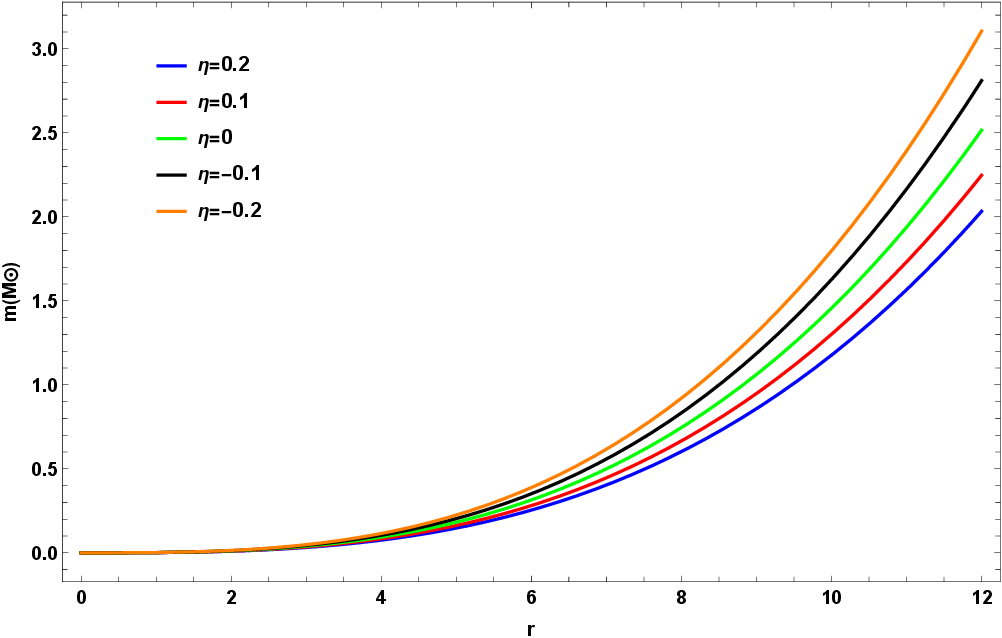}
\caption{Mass dependency on radial co-ordinate for CFL phase, $ m_{s} = 0, \triangle = 350MeV $}\label{7}
\end{figure}\\
\begin{figure}[ht!]
\centering
\includegraphics[scale=0.5]{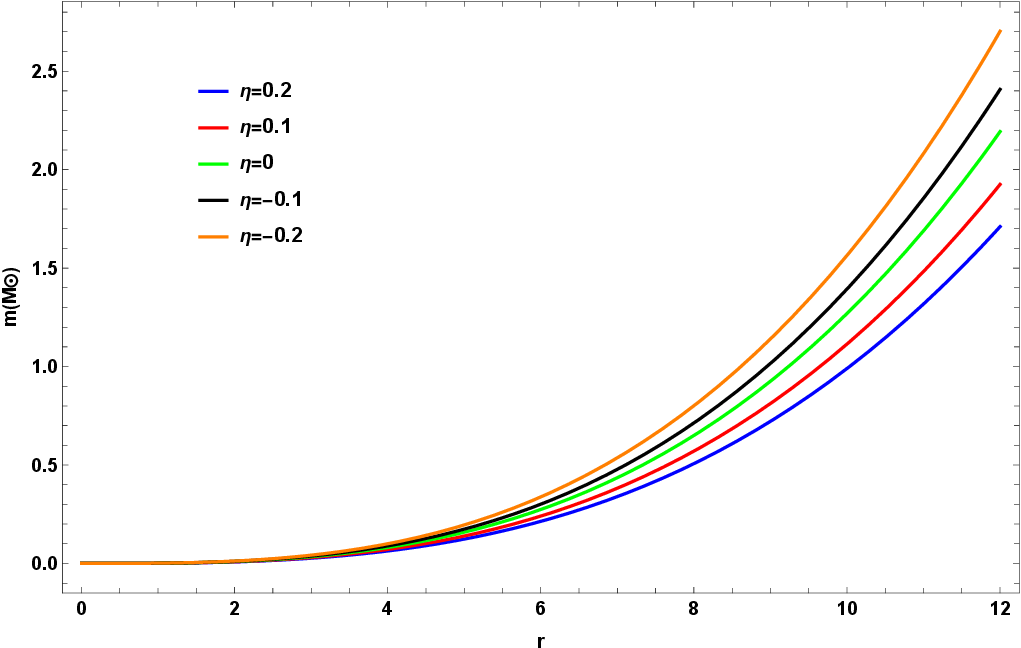}
\caption{Mass dependency on radial co-ordinate for CFL phase, $ m_{s} = 100 MeV, \triangle = 350MeV $}\label{8}
\end{figure}\\

\section{Stability analysis}\label{sec5}

\hspace{0.8cm}Alongside M-R curves, mass profiles, which are crucial for stellar framework, the stability analysis also needs to be discussed for the physical viability of our model. Thus we have done redshift analyss and examined the adiabatic index for the compact star configurations here.\\

\subsection{ Surface redshift }

\hspace{0.5cm}Redshift is a very important aspect of astrophysics as it enables a simplified investigation of the galaxy and cosmic properties. This occurs when an object's electromagnetic radiation shift towards the longer wavelengths in the spectrum. In astronomy, this term redshift refers to the measurable fractional change in wavelength between the emitted and received light, to indicate the prior events rather than the duration interval they occurred at. A photon leaving the centre towards the surface encounters more dispersion with energy loss. It is tied to the compactness of the stellar framework and thus important for describing stability. It connects interior configuration with the EoS. We have looked here into the isotropic, static and spherically symmetric structures with perfect fluid describing matter. We have the compactification factor as\\
\begin{equation}\label{25}
K(r) = \frac{m(r)}{r}.
\end{equation}\\
We have the surface redshift in terms of $ K $ as\\
\begin{equation}\label{26}
\textsl{Z} = \frac{1}{\sqrt{(1-2K)}} - 1.
\end{equation}\\
For isotropic setups, surface redshift does not exceed the value of 2 \cite{RT72}. Figure (\ref{9}) illustrates the change of $ \textsl{Z} $ with radial distance for the MIT bag model. Figures (\ref{10}) and (\ref{11}) show similar trends for the CFL phase EoS where $ \gamma = 0 $ represents the GR case. Figures  (\ref{9}), (\ref{10}) and (\ref{11}) show how $ \gamma $ influences the redshift parameter for the proposed models here. For all the cases it increase with the radial distance, it rises and peak towards the surface but stays within the limit of 2. Evidently, the redshift values here meet the stability criteria in this $f(R,L_{m},T)$ gravitational theory.\\ 
\begin{figure}[ht!]
\centering
\includegraphics[scale=0.5]{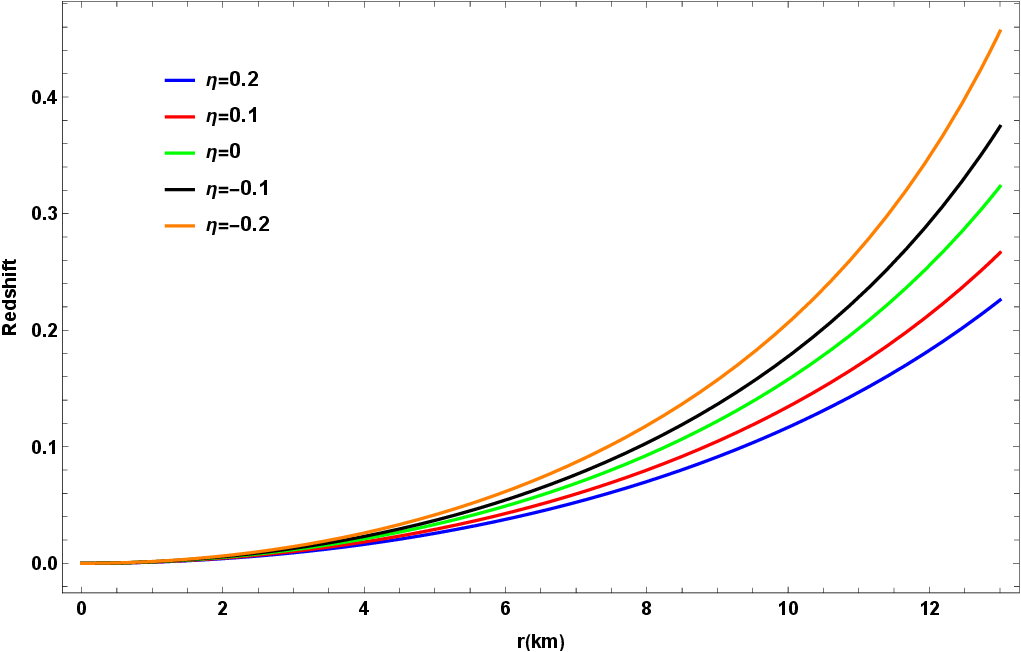}
\caption{For the MIT bag model, the nature of the redshift parameter, $ B = (168 MeV)^{4}  $}\label{9}
\end{figure}\\
\begin{figure}[ht!]
\centering
\includegraphics[scale=0.5]{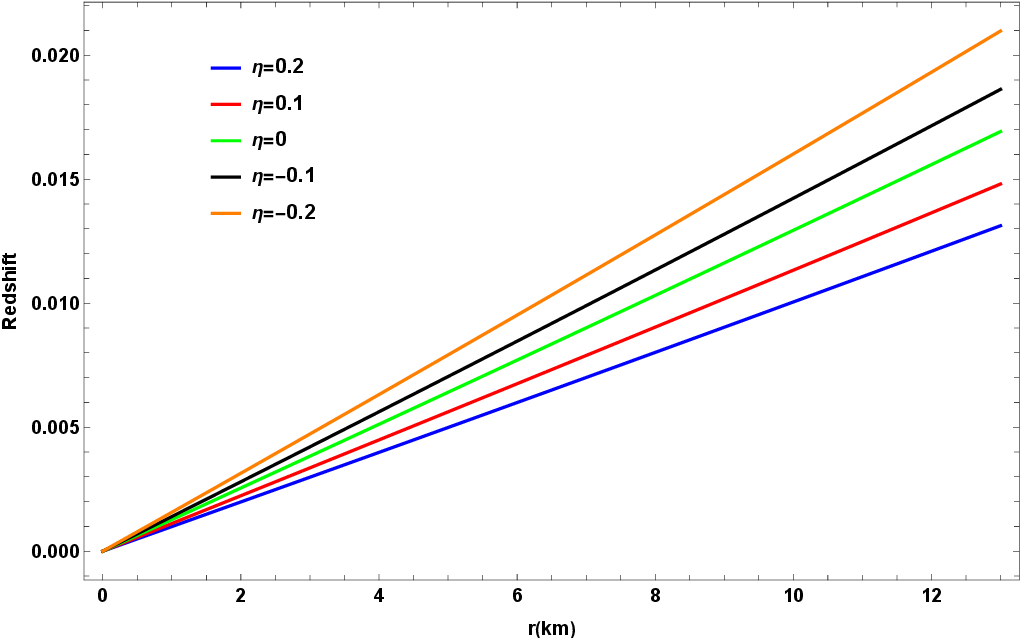}
\caption{For the CFL phase, the nature of the redshift parameter, $ m_{s} = 100 MeV, \triangle = 350MeV $}\label{10}
\end{figure}\\
\begin{figure}[ht!]
\centering
\includegraphics[scale=0.5]{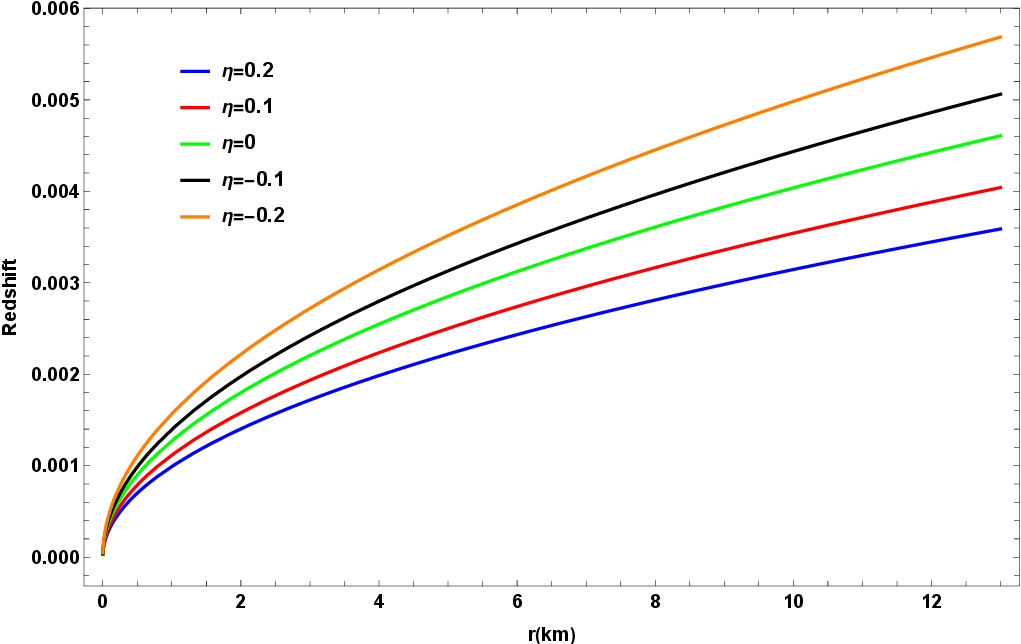}
\caption{For the CFL phase, the nature of the redshift parameter, $ m_{s} = 0, \triangle = 350MeV $}\label{11}
\end{figure}\\

\subsection{ Adiabatic index }

\hspace{0.5cm}Adiabatic index plays an important role for determining the stability region of isotropic fluid spheres. It was suggested by Chandrasekhar for spherically symmetric spacetime with perfect fluids in the works of \cite{RT75}. It results from the radial perturbations. Stability here is ensured for the real fundamental mode of radial oscillations \cite{RT77}. For the adiabatic index study, $ \Gamma > \frac{4}{3} $ gurantees the stability of an isotropic sphere, whereas lower values lead to instability. It is expressed as\\
\begin{equation}\label{27}
\Gamma = ( 1 + \frac{\rho}{p} )\frac{dp}{d\rho}.
\end{equation}\\
Figures (\ref{12}-\ref{14}) show adiabatic index plots for the MIT bag model and CFL phase EoS against radius. Clearly, we can see that $ \Gamma $ increases with radial co-ordinate and stays $ \Gamma > \frac{4}{3} $ within the interior implying the validity of our model. Ergoregion and non-linear instabilities might emerge for the ultra compact rotating stars. Prior consideration of static framework in our study avoids such instabilities.\\
\begin{figure}[ht!]
\centering
\includegraphics[scale=0.5]{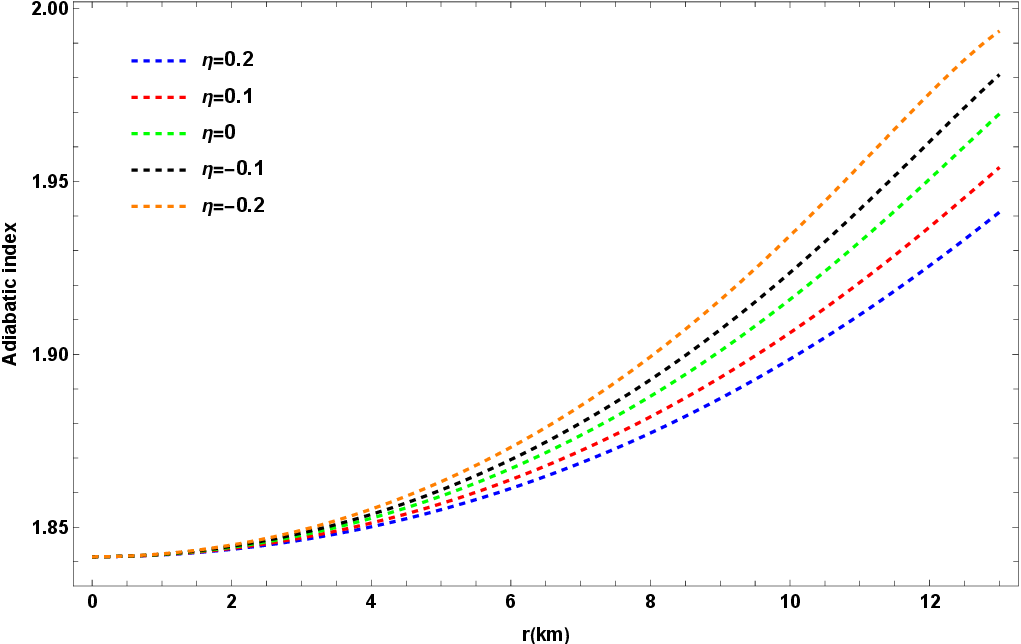}
\caption{Adaibatic index behavior against radius is shown for the MIT bag model}\label{12}
\end{figure}\\
\begin{figure}[ht!]
\centering
\includegraphics[scale=0.5]{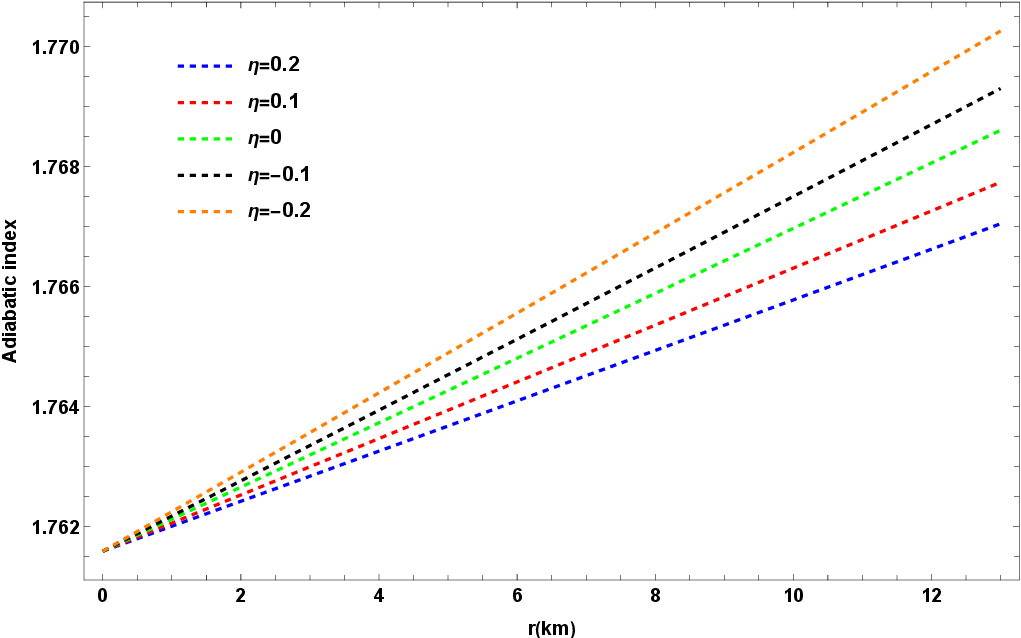}
\caption{Adaibatic index behavior against radius is shown for CFL phase with $ m_{s} = 100 MeV $}\label{13}
\end{figure}\\
\begin{figure}[ht!]
\centering
\includegraphics[scale=0.5]{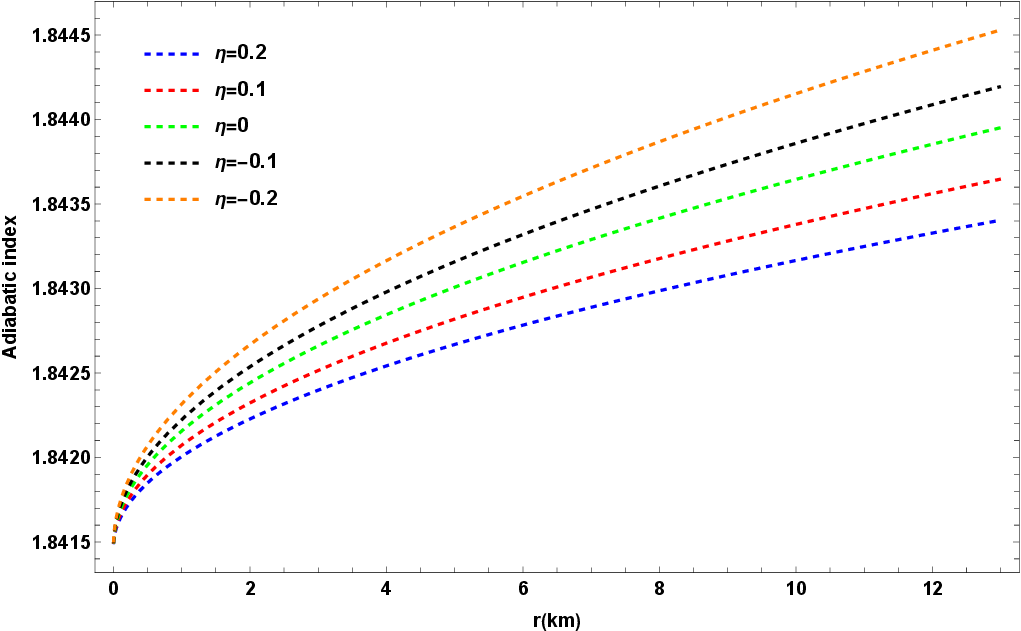}
\caption{Adaibatic index behavior against radius is shown for CFL phase with $ m_{s} = 0 $}\label{14}
\end{figure}\\

\section{Discussions and conclusion}\label{sec6}

\hspace{0.8cm}In this paper, we have delved into the characteristic features of the strange stars in modified $f(R,L_{m},T)$ gravitational theory. This study utilizes the model of $ f(R,L_{m},T) = R + \gamma TL_{m} $. This model provides a new look at the compact star's internal structure enabling exploration of modified gravity effects on star properties. Different stellar structures from the modified TOV equations with the MIT bag and CFL phase EoS have been obtained here. In this work, we have made the calculations for the GWE frequencies for these EoSs in $f(R,L_{m},T)$ gravitational framework. With the parameter lying in $ \gamma \epsilon [-0.2,0.2] \epsilon_{1} $ where $ \epsilon_{1} = 10^{-79}s^{4}/kg^{2} $, the M-R curves with MIT bag and CFL phase show good match with the observational data. A significant point from this research is that compact stellar structures with these EoSs can echo GWs. Using redshift analysis and adiabatic index study we have also discussed the model stability and thus our model appears viable.\\
GW frequencies can be instrumental in figuring out the properties of various compact stars. Detection of this frequency experimentally will help reveal the star's internal composition along with its physical features. The echo frequencies are derived here on the theoretical grounds. Experimentally detectable frequency predictions will be greatly useful here. Our obatained range of GWEs is in the range of $ 7.5-11 $ kHz which is  within range of the advanced LIGO, VIRGO and KAGRA detectors which are aimed at GW frequencies $ \backsim 20Hz - 4 kHz $, amplitude $ \backsim 2 \times 10^{-22} - 4 \times 10^{-24} strain/\sqrt{Hz} $ \cite{RT80,RT82}. At present, observatories are running with sensitivity of $ \geq 2 \times 10^{-23} strain/\sqrt{Hz} $ operating at $ 3 kHz $ \cite{RT85}. So detection of such frequencies is very much promising for uncovering most of the compact star's internal compositions. We can thus conclude from the above outcomes that $f(R,L_{m},T)$ gravity enables the formation and stability of strange stars explaining the observed candidates. Echo frequencies for such structures are also provided by us in this work. Future GW echo data from these candidates could help pick the best EoSs or constrain them enhancing the understanding of these star's properties. As a result, the underlying science of the high regime would be clearly resolved in the upcoming future.\\

\end{document}